\input harvmac
\input epsf

\def\Title#1#2{\rightline{#1}\ifx\answ\bigans\nopagenumbers\pageno0\vskip1in
\else\pageno1\vskip.8in\fi \centerline{\titlefont #2}\vskip .5in}
%

%
%
\ifx\epsfbox\UnDeFiNeD\message{(NO epsf.tex, FIGURES WILL BE IGNORED)}
\def\figin#1{\vskip2in}
\else\message{(FIGURES WILL BE INCLUDED)}\def\figin#1{#1}
\fi
\def\Fig#1{Fig.~\the\figno\xdef#1{Fig.~\the\figno}\global\advance\figno
 by1}
%
%
%
%
\def\Ifig#1#2#3#4{
\goodbreak\midinsert
\figin{\centerline{\epsfysize=#4truein\epsfbox{#3}}}
\narrower\narrower\noindent{\footnotefont
{\bf #1:}  #2\par}
\endinsert
}

%
%
\font\ticp=cmcsc10

\def\sq{{\vbox {\hrule height 0.6pt\hbox{\vrule width 0.6pt\hskip 3pt
   \vbox{\vskip 6pt}\hskip 3pt \vrule width 0.6pt}\hrule height 0.6pt}}}

\def\ajou#1&#2(#3){\ \sl#1\bf#2\rm(19#3)}
\def\jou#1&#2(#3){,\ \sl#1\bf#2\rm(19#3)}

\def\frac#1#2{{#1\over#2}}

\def\eg{{\it e.g.}}

\def\calo{{\cal O}}

\def\hatT{{\hat T}}
\def\bx{{\bf x}}
\def\bp{{\bf p}}
\def\by{{\bf x}}
\def\bq{{\bf p}}
\def\hatx{{\hat x}}

\def\hatbx{{\hat {\bf x}}}

\def\bx{{\bf x}}
\def\bp{{\bf p}}
\def\by{{\bf y}}
\def\bq{{\bf q}}
\def\bk{{\bf k}}
\overfullrule=0pt
%
%
%
\lref\Bankslittle{
T.~Banks,
``Cosmological breaking of supersymmetry or little Lambda goes back to  the future. II,''
arXiv:hep-th/0007146.
}
\lref\Fischler{W. Fischler, unpublished (2000)\semi
W. Fischler, ``Taking de Sitter seriously," Talk given at
Role of Scaling Laws in Physics and Biology (Celebrating
the 60th Birthday of Geoffrey West), Santa Fe, Dec. 2000.}
\lref\tHooftGX{
G.~'t Hooft,
``Dimensional Reduction In Quantum Gravity,''
arXiv:gr-qc/9310026.
}
\lref\BanksBY{
T.~Banks, L.~Susskind and M.~E.~Peskin,
``Difficulties For The Evolution Of Pure States Into Mixed States,''
Nucl.\ Phys.\ B {\bf 244}, 125 (1984).
}
\lref\SusskindVU{
L.~Susskind,
``The World as a hologram,''
J.\ Math.\ Phys.\  {\bf 36}, 6377 (1995)
[arXiv:hep-th/9409089].
}
\lref\MaldacenaRE{
J.~M.~Maldacena,
``The large N limit of superconformal field theories and supergravity,''
Adv.\ Theor.\ Math.\ Phys.\  {\bf 2}, 231 (1998)
[Int.\ J.\ Theor.\ Phys.\  {\bf 38}, 1113 (1999)]
[arXiv:hep-th/9711200].
}
\lref\FreivogelEX{
B.~Freivogel, S.~B.~Giddings and M.~Lippert,
``Toward a theory of precursors,''
Phys.\ Rev.\ D {\bf 66}, 106002 (2002)
[arXiv:hep-th/0207083].
}
\lref\GiddingsJQ{
S.~B.~Giddings,
``Flat-space scattering and bulk locality in the AdS/CFT  correspondence,''
Phys.\ Rev.\ D {\bf 61}, 106008 (2000)
[arXiv:hep-th/9907129].
}
\lref\SusskindVK{
L.~Susskind,
``Holography in the flat space limit,''
arXiv:hep-th/9901079.
}
\lref\PolchinskiRY{
J.~Polchinski,
``S-matrices from AdS spacetime,''
arXiv:hep-th/9901076.
}
\lref\SusskindIF{
L.~Susskind, L.~Thorlacius and J.~Uglum,
``The Stretched horizon and black hole complementarity,''
Phys.\ Rev.\ D {\bf 48}, 3743 (1993)
[arXiv:hep-th/9306069].
}
\lref\GiddingsPT{
S.~B.~Giddings and M.~Lippert,
``Precursors, black holes, and a locality bound,''
Phys.\ Rev.\ D {\bf 65}, 024006 (2002)
[arXiv:hep-th/0103231].
}
\lref\GiddingsQT{
S.~B.~Giddings,
``Why aren't black holes infinitely produced?,''
Phys.\ Rev.\ D {\bf 51}, 6860 (1995)
[arXiv:hep-th/9412159].
}
\lref\ChristensenJC{
S.~M.~Christensen and S.~A.~Fulling,
``Trace Anomalies And The Hawking Effect,''
Phys.\ Rev.\ D {\bf 15}, 2088 (1977).
}
\lref\StromingerTN{
A.~Strominger,
``Les Houches lectures on black holes,''
arXiv:hep-th/9501071.
}
\lref\GiddingsPJ{
S.~B.~Giddings,
``Quantum mechanics of black holes,''
arXiv:hep-th/9412138, in {\sl High energy physics and cosmology 1994}, proceedings of the 1994 Summer school in  high energy physics and cosmology, E. Gava, A. Masiero, K.S. Narain, S. Randjbar-Daemi, Q. Shafi, eds. (ICTP Series in Theoretical Physics, v. 11, World Scientific, 1995, River Edge, N.J.).
}
\lref\BoussoNF{
R.~Bousso,
``Positive vacuum energy and the N-bound,''
JHEP {\bf 0011}, 038 (2000)
[arXiv:hep-th/0010252].
}
\lref\GibbonsMU{
G.~W.~Gibbons and S.~W.~Hawking,
``Cosmological Event Horizons, Thermodynamics, And Particle Creation,''
Phys.\ Rev.\ D {\bf 15}, 2738 (1977).
}
\lref\DanielssonWB{
U.~H.~Danielsson and M.~E.~Olsson,
``On thermalization in de Sitter space,''
arXiv:hep-th/0309163.
}
\lref\BoussoNF{
R.~Bousso,
``Positive vacuum energy and the N-bound,''
JHEP {\bf 0011}, 038 (2000)
[arXiv:hep-th/0010252].
}
\lref\DanielssonTD{
U.~H.~Danielsson, D.~Domert and M.~Olsson,
``Miracles and complementarity in de Sitter space,''
Phys.\ Rev.\ D {\bf 68}, 083508 (2003)
[arXiv:hep-th/0210198].
}
\lref\DysonNT{
L.~Dyson, J.~Lindesay and L.~Susskind,
``Is there really a de Sitter/CFT duality,''
JHEP {\bf 0208}, 045 (2002)
[arXiv:hep-th/0202163].
}
\lref\HawkingRA{
S.~W.~Hawking,
``Breakdown Of Predictability In Gravitational Collapse,''
Phys.\ Rev.\ D {\bf 14}, 2460 (1976).
}
\lref\HawkingSW{
S.~W.~Hawking,
``Particle Creation By Black Holes,''
Commun.\ Math.\ Phys.\  {\bf 43}, 199 (1975).
}
\lref\GHM{S.~B.~Giddings, J.~Hartle, and D.~Marolf, work in progress.}
\lref\GiddingsZW{
S.~B.~Giddings,
``The fate of four dimensions,''
Phys.\ Rev.\ D {\bf 68}, 026006 (2003)
[arXiv:hep-th/0303031].
}
\lref\GKP{
S.~B.~Giddings, S.~Kachru and J.~Polchinski,
``Hierarchies from fluxes in string compactifications,''
Phys.\ Rev.\ D {\bf 66}, 106006 (2002)
[arXiv:hep-th/0105097].
}
\lref\tHooftRE{
G.~'t Hooft,
``On The Quantum Structure Of A Black Hole,''
Nucl.\ Phys.\ B {\bf 256}, 727 (1985).
}
\lref\tHooftUN{
G.~'t Hooft,
``Horizon operator approach to black hole quantization,''
arXiv:gr-qc/9402037.
}
\lref\StephensAN{
C.~R.~Stephens, G.~'t Hooft and B.~F.~Whiting,
``Black hole evaporation without information loss,''
Class.\ Quant.\ Grav.\  {\bf 11}, 621 (1994)
[arXiv:gr-qc/9310006].
}
\lref\tHooftFN{
G.~'t Hooft,
 ``Fundamental Aspects Of Quantum Theory Related To The Problem Of Quantizing
Black Holes,''
{\it Prepared for Fundamental Aspects of Quantum Theory: Conference on Foundations of Quantum Mechanics to Celebrate 30 Years of the Aharonov-
Bohm Effect (AB Conference), Columbia, South Carolina, 14-16 Dec 1989}.
}
\lref\VerlindeSG{
E.~Verlinde and H.~Verlinde,
``A Unitary S matrix and 2-D black hole formation and evaporation,''
Nucl.\ Phys.\ B {\bf 406}, 43 (1993)
[arXiv:hep-th/9302022].
}
\lref\SchoutensST{
K.~Schoutens, H.~Verlinde and E.~Verlinde,
``Black hole evaporation and quantum gravity,''
arXiv:hep-th/9401081.
}
\lref\KiemIY{
Y.~Kiem, H.~Verlinde and E.~Verlinde,
``Black hole horizons and complementarity,''
Phys.\ Rev.\ D {\bf 52}, 7053 (1995)
[arXiv:hep-th/9502074].
}
\lref\CallanRS{
C.~G.~Callan, S.~B.~Giddings, J.~A.~Harvey and A.~Strominger,
``Evanescent Black Holes,''
Phys.\ Rev.\ D {\bf 45}, 1005 (1992)
[arXiv:hep-th/9111056].
}
\lref\BanksVP{
T.~Banks,
 ``A critique of pure string theory: Heterodox opinions of diverse
dimensions,''
arXiv:hep-th/0306074.
}
\lref\BanksYP{
T.~Banks and W.~Fischler,
``M-theory observables for cosmological space-times,''
arXiv:hep-th/0102077.
}
\lref\tHooftEK{
G.~'t Hooft,
``Horizons,''
arXiv:gr-qc/0401027.
}
\lref\LoweAC{
D.~A.~Lowe, J.~Polchinski, L.~Susskind, L.~Thorlacius and J.~Uglum,
``Black hole complementarity versus locality,''
Phys.\ Rev.\ D {\bf 52}, 6997 (1995)
[arXiv:hep-th/9506138].
}
\lref\EardleyRE{
D.~M.~Eardley and S.~B.~Giddings,
``Classical black hole production in high-energy collisions,''
Phys.\ Rev.\ D {\bf 66}, 044011 (2002)
[arXiv:gr-qc/0201034].
}
\lref\AichelburgDH{
P.~C.~Aichelburg and R.~U.~Sexl,
``On The Gravitational Field Of A Massless Particle,''
Gen.\ Rel.\ Grav.\  {\bf 2}, 303 (1971).
}
\lref\tHooftFR{
G.~'t Hooft,
``The Black Hole Interpretation Of String Theory,''
Nucl.\ Phys.\ B {\bf 335}, 138 (1990).
}
\lref\CopelandET{
E.~J.~Copeland, A.~R.~Liddle and D.~Wands,
``Exponential potentials and cosmological scaling solutions,''
Phys.\ Rev.\ D {\bf 57}, 4686 (1998)
[arXiv:gr-qc/9711068].
}
\lref\RatraRM{
B.~Ratra and P.~J.~Peebles,
``Cosmological Consequences Of A Rolling Homogeneous Scalar Field,''
Phys.\ Rev.\ D {\bf 37}, 3406 (1988).
}
\lref\KachruAW{
S.~Kachru, R.~Kallosh, A.~Linde and S.~P.~Trivedi,
``De Sitter vacua in string theory,''
Phys.\ Rev.\ D {\bf 68}, 046005 (2003)
[arXiv:hep-th/0301240].
}
\lref\GiddingsPT{
S.~B.~Giddings and M.~Lippert,
``Precursors, black holes, and a locality bound,''
Phys.\ Rev.\ D {\bf 65}, 024006 (2002)
[arXiv:hep-th/0103231].
}
\lref\GiddingsZW{
S.~B.~Giddings,
``The fate of four dimensions,''
Phys.\ Rev.\ D {\bf 68}, 026006 (2003)
[arXiv:hep-th/0303031].
}
\lref\Thorne{
K.~S.~Thorne,
``Nonspherical Gravitational Collapse: A Short Review,''
in {\it Magic Without Magic}, ed.~J.~R.~Klauder, San Francisco 1972, 231-258.
}
\lref\HorowitzHE{
G.~T.~Horowitz and J.~Maldacena,
``The black hole final state,''
arXiv:hep-th/0310281.
}
\lref\FiolaIR{
T.~M.~Fiola, J.~Preskill, A.~Strominger and S.~P.~Trivedi,
``Black Hole Thermodynamics And Information Loss In Two-Dimensions,''
Phys.\ Rev.\ D {\bf 50}, 3987 (1994)
[arXiv:hep-th/9403137].
}
\lref\Page{
D.~N.~Page,
``Information in black hole radiation,''
Phys.\ Rev.\ Lett.\  {\bf 71}, 3743 (1993)
[arXiv:hep-th/9306083].
}
\lref\HartleTP{
J.~B.~Hartle and S.~W.~Hawking,
``Path Integral Derivation Of Black Hole Radiance,''
Phys.\ Rev.\ D {\bf 13}, 2188 (1976).
}
\lref\YoshinoTX{
H.~Yoshino and Y.~Nambu,
``Black hole formation in the grazing collision of high-energy particles,''
Phys.\ Rev.\ D {\bf 67}, 024009 (2003)
[arXiv:gr-qc/0209003].
}
\lref\JacobsonVX{
T.~Jacobson,
``Introduction to quantum fields in curved spacetime and the Hawking effect,''
arXiv:gr-qc/0308048.
}
\lref\DysonNT{
L.~Dyson, J.~Lindesay and L.~Susskind,
``Is there really a de Sitter/CFT duality,''
JHEP {\bf 0208}, 045 (2002)
[arXiv:hep-th/0202163].
} 
\lref\BanksYP{
T.~Banks and W.~Fischler,
``M-theory observables for cosmological space-times,''
arXiv:hep-th/0102077.
}
\lref\SusskindIF{
L.~Susskind, L.~Thorlacius and J.~Uglum,
``The Stretched horizon and black hole complementarity,''
Phys.\ Rev.\ D {\bf 48}, 3743 (1993)
[arXiv:hep-th/9306069].
}
\lref\PolchinskiTA{
J.~Polchinski,
``String theory and black hole complementarity,''
arXiv:hep-th/9507094.
}
\lref\ParikhPY{
M.~K.~Parikh, I.~Savonije and E.~Verlinde,
Phys.\ Rev.\ D {\bf 67}, 064005 (2003)
[arXiv:hep-th/0209120].
}
\Title{\vbox{\baselineskip12pt
\hbox{hep-th/0402073}
}}
{\vbox{\centerline{The information paradox and the locality bound}
}}
\centerline{{\ticp Steven B. Giddings}\footnote{$^\dagger$}
{Email address:
giddings@physics.ucsb.edu}  and {\ticp Matthew Lippert}\footnote{$^\star$}
{Email address:
lippert@physics.ucsb.edu} }
\bigskip\centerline{ {\sl Department of Physics}}
\centerline{\sl University of California}
\centerline{\sl Santa Barbara, CA 93106}

\bigskip\bigskip
\centerline{\bf Abstract}
Hawking's argument for information loss in black hole evaporation rests on the assumption of independent Hilbert spaces for the interior and exterior of a black hole.  We argue that such independence cannot be established without incorporating strong gravitational effects that undermine locality and invalidate the use of quantum field theory in a semiclassical background geometry.  These considerations should also play a role in a deeper understanding of horizon complementarity.

\Date{}

\newsec{Introduction}

Hawking's discovery of black hole radiance\refs{\HawkingSW} initiated a crisis in theoretical physics, which  crystallized in Hawking's 1976 paper\refs{\HawkingRA} arguing for breakdown of quantum mechanics in black hole evaporation.  This crisis is the black hole information paradox.\foot{For reviews, see \refs{\GiddingsPJ,\StromingerTN} .}  

In short, locality implies that information that falls into a big black hole cannot escape until the final stages of its evaporation.   But then the remaining energy in the black hole is too small to radiate the information except on extremely long timescales.  Therefore either the information has been fundamentally destroyed, as Hawking advocated, or a long-lived black hole remnant is left behind.  There are very general arguments that remnants, which would necessarily have infinite internal states, would be infinitely produced in everyday physical processes (see \GiddingsQT\ and references therein).  That leaves information destruction, which is believed to be very insidious and dangerous.   Once allowed in physics, through virtual processes it apparently contaminates all of physics, with the unacceptable result\refs{\BanksBY} that all physical processes would appear coupled to a heat bath at the Planck temperature.  This is the essence of the paradox.

This crises led to the radical proposal\refs{\tHooftGX,\SusskindVU} that physics is at a deep level not local but rather holographic.  In particular, it is now widely believed that quantum mechanics is saved through a non-local escape of information in black hole evaporation.  Aspects of this holographic picture are believed to be manifest in string theory in the conjectured AdS/CFT correspondence\refs{\MaldacenaRE}, although the foundations of holography and, in particular, the question of how to reconsctruct approximately local bulk physics from holographic data remains a mystery\refs{\GiddingsJQ}.\foot{Though see \refs{\SusskindVK,\PolchinskiRY} for attempts.}

While these beliefs suggest an escape from the paradox, the picture is incomplete.   One element that is missing is a clear statement of where Hawking's original calculation\refs{\HawkingRA} breaks down; so far the loophole through which nonlocality manifests itself has not been sharply identified.  

Another missing element is a reconciliation of the descriptions of physics as seen by observers who stay outside the black hole and those who fall in.  The relationship between these descriptions is the subject of  {\it black hole complementarity}\refs{\SusskindIF,\StephensAN}, which states that there is no way to compare observations inside and outside the horizon to find a conflict with the the ban on quantum xeroxing of information.  One might take this one step further, and conjecture that complementarity is realized through the statement that observables inside and outside the black hole cannot be simultaneously described, much as $x$ and $p$ cannot be simultaneously measured in quantum mechanics. These ideas have been extended to discussion of cosmological horizons\refs{\Bankslittle,\BoussoNF,\DysonNT}, where they may imply that the physics of a de Sitter universe can be encoded in the dynamics of a single causal patch\refs{\Bankslittle,\Fischler,\ParikhPY}.  However, we are still left with many questions about the precise rules of complementarity; for example, to which situations does it apply, and how does it constrain and relate degrees of freedom?

This paper is an attempt to address the former question, while taking a small step towards better understanding complementarity.  In particular, it is clear that in a theory with dynamical quantum gravity, whether string theory or some other formulation, the concept of locality is approximate and there are contexts where it should fail due to strong gravitational or other dynamics.  In the next section, we revisit and expand on a general criterion for such contexts, first proposed in \refs{\GiddingsPT}, namely the {\it locality bound}.   In section three, we then apply locality bound arguments to Hawking's calculation and suggest that self-consistency of the argument that information does not escape a black hole relies on assuming that physics is local in regimes where the locality bound would indicate it should not be.  This identifies a potential flaw in Hawking's original reasoning, and hence potentially a  resolution of the black hole information paradox.  At the same time, the locality bound appears to provide deeper rationale for the notion of black hole and more generally horizon complementarity.  There is a long history of discussions of the relevance of ultrahigh blueshifts near black hole horizons\refs{\tHooftRE\tHooftFN\tHooftFR\StephensAN\tHooftUN\VerlindeSG\SchoutensST\KiemIY-\LoweAC}; this paper works towards providing a sharper argument against information loss and outlines an approach to a deeper understanding of complementarity that could provide criteria for its application.
We close with a discussion of some other aspects of the locality bound, particularly its relevance to cosmology, the issue of formulating it more precisely, and the question of a more fundamental formulation of physics that respects its inherent constraints on dynamics.

\newsec{Bounding local physics}

In ordinary quantum field theory with observeables generically denoted $\calo_i(x)$, locality is encoded in the statement that
\eqn\localdef{[\calo_i(x),\calo_j(y)]=0}
for all spacelike separated pairs of spacetime points $x$ and $y$ and all labels $i$ and $j$.  For example, in the simplest case of a scalar field, the observables include the field $\phi(x)$ and its derivatives.  As we will review in the next section, this statement plays a central role in arguing for loss of information into a black hole.  We can think of this statement as saying we can excite independent quantum degrees of freedom at any spacelike separated points.

When we think of how locality should be stated within the context of the theory of the world, in particular incorporating dynamical gravity, \localdef\ relies on some idealizations.  First, this statement is made within the context of quantum fields in a {\it fixed} background spacetime -- it ignores dynamical gravitational effects.  Secondly, this statement refers to field operators at a point.  These operators are a superposition of creation and annihilation operators for particles of all momenta and energies up to infinity.  Even in contexts where gravity is negligible, for example in describing the physics of our experiments at accelerators, we clearly don't deal directly with such idealized objects.

Indeed, the latter idealization is addressed by working with {\it wavepackets} -- operators with essentially finite spread in position and momentum.  Any experiments that we perform use such constructs.  While there are many ways of parametrizing such wavepackets, gaussian wavepackets are particularly simple.  For example with scalars in $d+1$-dimensional flat space, we might consider operators of the form
\eqn\gaussop{\phi_\delta(\bx,\bp,t) = \int d^d  \bx'   {e^{-(\bx'-\bx)^2/2\delta^2
- i\bp\cdot(\bx'-\bx)} \over  
(\sqrt{2\pi}\delta)^d}  \phi(\bx', t) \ .}
This operator creates or annihilates a particle at time $t$ with approximate position $\bx$ and momentum $\bp$, and with spreads of $\delta$ and $1/\delta$ respectively.\foot{One could likewise explicitly incorporate a spread in time and energy, although such a spread can be approximately subsumed in the spread of position and momentum.}  When we consider a state created at an accelerator, it is one of this form, with a nearly definite momentum and location for the particles.

Since operators like \gaussop\ are more physically realistic, it makes sense to restate locality in terms of them.  The basic idea is simple: since these operators create gaussian wavepackets with finite spread in position and momenta, \localdef\ should be replaced by an expression that is non-vanishing but exponentially small.  

In particular, consider the equal-time commutator of two wavepackets of the form \gaussop.
Using the equal-time canonical commutation relations and integrating the remaining gaussian integral, we find
\eqn\approxlocal{[\phi_\delta(\bx,\bp,t), {\dot \phi}_\delta(\by,\bq,t)] =  i {1\over 
({2\sqrt{\pi}}\delta)^d} e^{-(\bx-\by)^2/4\delta^2 -(\bp+\bq)^2 \delta^2/4 + i (\bp-\bq)(\bx-\by)/2} . }
Note that \approxlocal\ assumes the canonical value in the limit $\delta \to 0$ and $\phi_\delta(\bx,\bp,t) \to \phi(\bx,t)$.  We  therefore restate locality as the condition 
\eqn\localdefwp{|[\calo_\delta(\bx,\bp,t),\calo_\delta(\by,\bq,t)]| \roughly< {e^{-(\bx-\by)^2/4\delta^2} \over 
({2\sqrt{\pi}}\delta)^d}\ .}
%

%
%

Overcoming the first idealization is more subtle.  Indeed, we do not yet understand a complete formulation of approximately local operators within the context of a quantum theory of gravity, although one can take some very concrete steps towards overcoming various objections to the existence of such operators\refs{\GHM}.  (In string theory we might expect such operators to be appropriate combinations of string field operators.)  Moreover, we know that in a complete theory of quantum gravity, there must be operators that reduce to standard field theory  operators in the approximation where gravitational effects are small.  Once such a theory is eventually fully understood, the weak-gravity limit of such operators would be used, for example, to describe creating a particle at an approximately definite position and momentum, like we do in actual experiments at accelerators.

Given that in a full theory of quantum gravity there should exist operators that, in a background corresponding to a large semiclassical geometry, correspond to creating a particle with some approximate momentum and position, we can ask about the properties of such operators.  Clearly for low momenta and distant positions, statements such as \localdefwp\ should hold.  But eventually, as dynamical gravity becomes important, we have no right to trust such statements and indeed, expect them to break down.  Specifically, consider, in the framework of a theory of dynamical gravity, working in a background corresponding to Minkowski space and attempting to study the creation/annihiliation of two particles with approximate positions and momenta $(\bx,\bp)$ and $(\bx',\bp')$ respectively.  The very notion of a benign semiclassical background clearly fails when the mutual gravitational field of the two particles becomes strong.\foot{On the other hand, we believe that the gravitationally dressed operator that creates a single high-momentum particle in otherwise flat space is easily treated semiclassically, as it differs from a low-momentum particle by a boost.  The semiclassical geometry should correspond to the Aichelberg-Sexl solution\refs{\AichelburgDH}.}  This is the origin of the {\it locality bound}\refs{\GiddingsPT}:  specifically, we expect locality in field theory, as encoded in statements like  \approxlocal, to fail when the mutual gravitational backreaction of the states 
created/annihilated by the respective field operators becomes strong.

As an example, consider the simple case where we work in the center of mass frame, with particles of momenta $\bp$ and $-\bp$.  Clearly the gravitational field becomes strong when
\eqn\lbcrit{(\bx-\bx')^{d-2}\roughly< \vert\bp\vert\ .}
Indeed, according to Thorne's hoop conjecture\refs{\Thorne}, which has now been shown to be qualitatively correct in collisions of high-energy particles\refs{\EardleyRE, \YoshinoTX}, two such particles have a sufficiently violent effect on the geometry to form a black hole.  For such wavepackets, we expect expressions of locality such as \approxlocal\ to fail.

This condition for breakdown of locality due to strong gravity, and specifically black hole formation, should be generic to any theory that reproduces general relativity at longer distances.  In the context of a specific theory, there may be even tighter constraints.  For example, it is widely believed that string theory is the correct theory of quantum gravity.  In this case, production of long strings may lead to an even tighter locality bound, as discussed in \GiddingsPT.  (The locality bound was originally formulated to help understand the problem of precursors in the AdS/CFT correspondence; for more discussion see \FreivogelEX.) Which effect is dominant apparently depends on detailed dynamics.  In this paper we focus on the constraints arising from strong gravity, bearing in mind these arguments may have obvious extensions in cases where other nonlocal effects are relevant.

\newsec{Arguing for and against information loss}

\subsec{Locality and information loss}

We begin by reviewing the arguments for information loss.  Since Hawking's original paper \HawkingRA, there have been a number of restatements of the basic logic.  However, they all essentially rely on field theory locality, \localdef.  Let us recall the reasoning.  

Suppose that we draw the Penrose diagram for an evaporating black hole, fig.~1.  Then there are spacelike slices like $S$ that cut across any matter that falls into the black hole, and through the outgoing Hawking radiation.  Since any point $x$ outside the horizon is spacelike separated from any point $y$ inside the horizon, field operators acting at these two points commute, \localdef.  This means that the state on any such spacelike slice $S$ can be decomposed into a state in a product of two independent Hilbert spaces, one inside the horizon and one outside.  In other words, the degrees of freedom inside the horizon and outside are completely independent.  The part of the state inside the black hole falls into the singularity and so in the final state, after the black hole has evaporated, must be traced over.  This produces a mixed state density matrix as the final state.  This basic argument has been run in many different guises, including in two-dimensional theories\refs{\CallanRS} where there is a great deal of analytical control, and with various choices for slices, see {\it e.g.} \refs{\FiolaIR}.
\Ifig{\Fig\bhPenrose}{The Penrose diagram of an evaporating black hole.  Spacelike slice $S$ passes through points $x$ and $y$, outside and inside the black hole, respectively.}{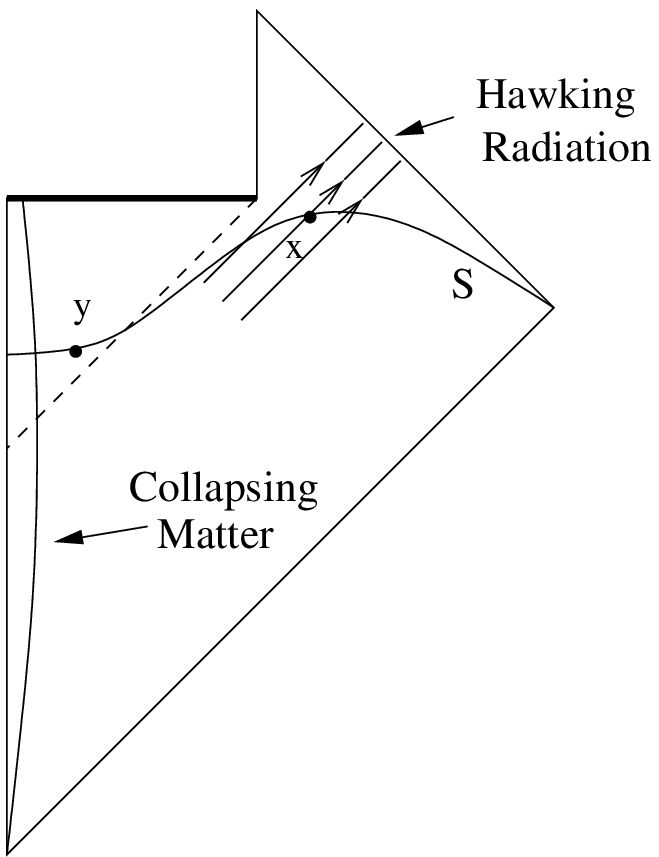}{3}

\subsec{The role of the locality bound}

Breakdown of the locality statement \localdef\ clearly undercuts the argument for information loss; in particular, a significant failure of locality removes the rationale for an independent  Hilbert space internal to the black hole.  Indeed, this is one interpretation of the idea of black hole complementarity\refs{\SusskindIF,\StephensAN}, that observations inside a black hole are not independent of those outside.   However, a clear derivation of black hole complementarity, and its role in Hawking's calculation of black hole radiance, has not yet been provided.  The locality bound suggests such a deeper rationale for this idea.

It is clear that, in accordance with the proposed locality bound, the dynamics of our ordinary experience is effectively local, since we have not yet directly encountered situations where the strong gravity condition \lbcrit\ holds.  The question is whether such a bound is relevant to black hole evaporation. We will argue the answer is yes.
To do so, we will follow the usual rules of quantum field theory in a background, used by Hawking in his seminal work\HawkingSW, and see where they come into conflict with the locality bound.

The basic idea is the following.  If information escapes a black hole of mass $M$ during its evaporation, it may escape at a relatively late time; indeed, Page\refs{\Page} has argued that it can escape as late as the time scale $\tau_{\rm evap}\sim M^3$ when the black hole has lost an appreciable part of its mass.\foot{Henceforth, for simplicity, we work in four dimensions.}  Thus, to compare information contained inside the black hole to that in the Hawking radiation, the slices $S$ of fig.~1 must be highly deformed.  In particular this means that there is a huge relative boost between the typical local observers at points $x$ and $y$.  We do not yet have a clear analog of the statement \lbcrit\ in a general spacetime, for generally separated points $x$ and $y$.  However, using the assumption that field theory is valid, we can relate the commutator $[\calo_i(x),\calo_j(y)]$ to a commutator on another spacelike slice, where  the criterion \lbcrit\ applies.  Because of the well known extreme blueshift encountered by tracing a late Hawking mode back to its origin near the horizon, the locality bound is found to be violated, indicating the breakdown of the argument for independence of the internal and external Hilbert spaces.

Specifically, imagine that an observer hovering in a spaceship outside a large black hole detects a Hawking particle of field $\phi$ at (approximate) position $x$ and momentum $p$, at some late time $\tau$.   We would like to know if this degree of freedom is independent of a degree of freedom measured by an observer that fell into the black hole early, around the time of its formation; for example consider an infalling particle of momentum $q$ at position $y$.  If these degrees of freedom are not independent, the Hilbert space does not factorize into internal and external Hilbert spaces, and the argument for information loss breaks down.
Observations on the field at $x$ and $y$ are described by the action of field operators of the form \gaussop\ on the Hilbert space of the field.

\Ifig{\Fig\RindlerRegion}{A Kruskal diagram for an eternal black hole.  The commutator between observations at points $x$ and $y$ is related by time evolution to that at points $x'$ and $y'$.  The dotted lines represent the boundary of the Rindler region described in the text.}{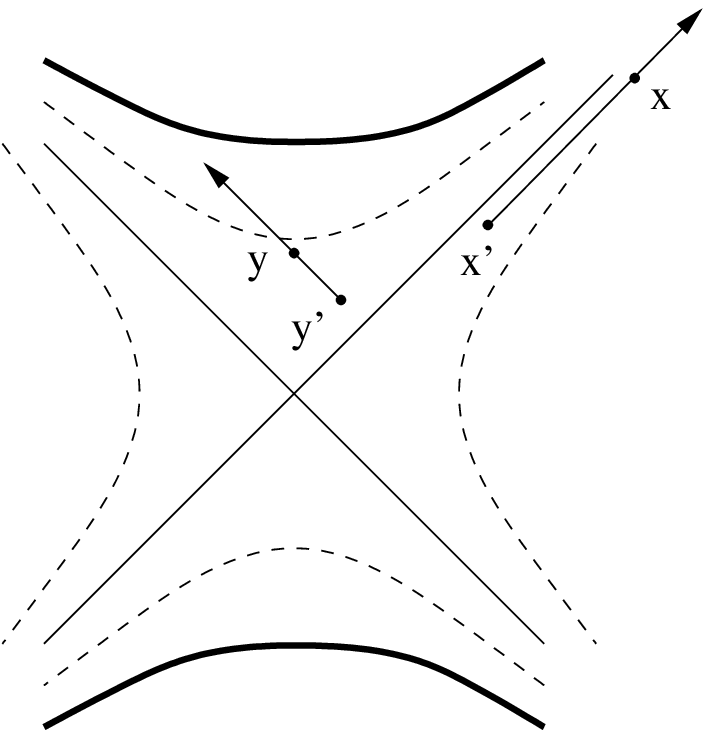}{4}

For any large $\tau$, the particles at $x$ and $y$ have a large relative boost.
 In order to simplify the kinematics of the problem, let us study the special case where both $x$ and $y$ lie in a region whose distance from the horizon is small as compared to the Schwarzschild radius of the black hole:
\eqn\rrdef{r-2M \ll 2M .}
This {\it Rindler region} is pictured in fig.~2; in this region the Schwarzschild metric 
\eqn\schmet{ds^2 = -\left(1-{2M\over r}\right)dt^2 + {dr^2\over \left(1-{2M\over r}\right)} + r^2 d\Omega^2} 
is well approximated by flat space in Rindler coordinates,
\eqn\rindler{ds^2=\pm\left(-\rho^2 d\theta^2 + d\rho^2\right) + dX_\perp^2\ }
where the $\pm$ refers to outside/inside the horizon.
This is easily seen to follow from the coordinate transformation
\eqn\rindxm{\rho=\sqrt{\pm8M(r-2M)}\ ,\ \theta=t/4M\ ,}
 together with defining natural local flat coordinates $X_\perp$ on the large two-sphere.
The transformation to the usual Minkowski metric 
\eqn\minkmet{ds^2=-dT^2 + dX^2 + dX_\perp^2}
is then
\eqn\minkxm{\eqalign{X^\pm&=T\pm X = \pm\rho e^{\pm\theta}\quad {\rm outside\ the\ horizon} \cr
X^\pm &= \rho e^{\pm\theta} \quad {\rm inside\ the\ horizon} \ .}}
%
These latter coordinates are limits of the Kruskal coordinates, and the usual Hartle-Hawking vacuum\refs{\HartleTP} for $\phi$ corresponds to the Minkowski vacuum.
Note that for a big black hole the Rindler region can be quite large, and the redshifts of this region relative to infinity quite mild.

The states created by the operators at $x$ and $y$ can now be investigated using flat space kinematics.  The role of the locality bound is most easily explored by boosting to the center of mass frame.  This is accomplished via the boost 
\eqn\boost{\theta\rightarrow\theta'=\theta-\tau/8M\ .}
This leaves the Rindler region invariant.

For illustration, consider classical massless particles originating at $x,y$ such that $(\rho,\theta) = (\rho_x,\tau/4M)$, $(\rho_y,0)$, corresponding to a Hawking particle emitted at some late time $\tau$ and a particle inside the black hole at a time around its formation.  Let them have Schwarzschild energies $E_x$ and $E_y$, respectively.  These correspond to  Minkowski momenta 
\eqn\cmmom{\eqalign{P_+' &= 0\ ,\ P_-'= {-4ME_x\over \rho_x} e^{\tau/8M}\cr
Q_+'&= {-4ME_y\over \rho_y} e^{\tau/8M}\ ,\ Q_-'=0\ }}
in the frame boosted by \boost.
(Due to the large boosts involved, we neglect mass terms.) The classical trajectories of the particles are
\eqn\traject{X_x^{'-}=-\rho_x e^{-\tau/8M}={\rm const.}\ ,\ X_y^{'+} = \rho_y e^{-\tau/8M}={\rm const.}\ }
For $E_x\sim E_y\sim E$ and $\rho_x\sim\rho_y\sim\rho$, these particles would form a black hole at Minkowski time 
\eqn\bhtime{\hatT'\sim {2ME\over \rho }e^{\tau/8M}\ .}
The proper size of this black hole is also of order $\hatT'$.  Therefore, for 
\eqn\bhtime{\tau\roughly> M\log(\rho/E), }
the black hole would have a transverse size comparable to the size of the horizon of the original black hole.  For $\tau\sim M^3$ its size is exponentially larger.

We are of course instead interested in the commutator of two observables, \eg\ 
\eqn\bhcom{[\phi_\delta(\bx,\bp,T'),{\dot \phi}_\delta(\by,\bq,T')]\ ,}  
by which we mean the corresponding expression in a full quantum theory of gravity.  Following Hawking, we suppose that this quantity can be well-approximated by the semiclassical result for QFT in a background, but then find that its vanishing conflicts with the locality bound.

Specifically, using methods of QFT in a gravitational background, we can rewrite \bhcom\ in terms of a commutator on the spacelike slice labeled by $\hatT'$ where the locality bound is saturated.  Formally, we can write
\eqn\commreln{[\phi_\delta(\bx,\bp,T'),{\dot \phi}_\delta(\by,\bq,T')] = [\calo_{\phi_\delta}(\hatT'), \calo_{\dot \phi_\delta}(\hatT')] }
where the operators on the RHS are just those which create a state at $\hatT'$ that evolves into the wavepackets at $T'$.  For operators $\phi_\delta$ which create a high-energy particle in a wavepacket, $\calo_{\phi_\delta}(\hatT')$ is well approximated by propagating that wavepacket classically backward in time.

This is most easily seen in the example of a scalar field that interacts only gravitationally.   In that case, we can use \gaussop\ and Green's theorem to propagate the wavepackets backward in time.  The calculation, given in appendix A, verifies that wavepackets propagate approximately classically,
\eqn\wppropback{\calo_{\phi_\delta}(\hatT') \sim {\tilde \phi}_\delta(\bx(\hatT'),\bp,\hatT') \ ,}
where the modified wavepacket operator ${\tilde \phi}_\delta$, given by (A.4),
is located at $\bx(\hatT') = \bx - (T'-\hatT') {\hat {\bf e}}_\bp$.
The RHS of \commreln\ is therefore well approximated by the commutator of these modified wavepackets
\eqn\commpropback{ [\phi_\delta(\bx,\bp,T'),{\dot \phi}_\delta(\by,\bq,T')] \sim [{\tilde \phi}_\delta(\bx(\hatT'),\bp,\hatT'),{\tilde {\dot \phi}}_\delta(\by(\hatT'),\bq,\hatT')] \ .}

If we were considering QFT in a background, this expression would vanish, up to the exponential tails discussed in the preceding section.  However,  we clearly see that the commutator at $T'$ is directly related to one at ${\hat T}'$ that  in the full theory has no right to vanish, by the locality bound.


In other words, while the commutator between observations carried out by an observer measuring the Hawking radiation and another falling into a black hole vanishes in the context of QFT in a background, computation of this commutator in the full theory implicitly makes reference to situations where the gravitational backreaction becomes strong and locality can no longer be trusted.  By referring back to the slice at $\hatT'$, we see that  what were presumed, using ordinary field theory, to be independent degrees of freedom, are not necessarily independent, because of the strong gravitational physics that arises when one tries to investigate their independence.  It is here that Hawking's calculation runs into the constraints of the locality bound.

Thus there is apparently no self-consistent description of black hole evaporation via QFT in a background which is valid throughout the entire region necessary to address the question of whether or not information escapes a black hole.  This apparently removes the rationale for the claim that black holes destroy information.

It's also worth  investigating how late in the black hole's evaporation a generic observation of a Hawking particle will violate the locality bound with a generic observation of an infalling observer.  We found that the size of the region of strong backreaction, where the internal and external degrees of freedom influence each other, is given by \bhtime\ in the center of mass frame.  The maximum impact parameter between the infalling particle and the outgoing Hawking particle is of order $M$, the black hole size.  So for $\rho$ some small fixed fraction of $M$ and for a typical $E\sim1/M$, we find 
\eqn\tlb{\tau_{lb} \sim M \log M^2\ ;}
any particle emerging in the Hawking radiation at a later time will violate the locality bound with a generic particle falling in during the formation of the black hole.  This timescale has previously appeared in the literature as the time that a particle or string falling into a black hole should take to spread over the horizon in the holographic picture\refs{\SusskindVU}.

\subsec{Discussion}

There has long been a sense that the ultra-high blueshift of a Hawking particle traced back to near the horizon should play an important role in the question of information loss.  In particular, in a large body of work 't Hooft and collaborators have  argued for the importance of this observation (see for example \refs{\tHooftRE\tHooftFR\tHooftFN\StephensAN-\tHooftUN}). This thread was also taken up by E. and H. Verlinde and collaborators\refs{\VerlindeSG\SchoutensST-\KiemIY} in the context of two-dimensional models for black hole formation and evaporation\CallanRS.

A central theme in all this work was that the interaction between the quanta can be treated in terms of large spatial shifts.  (Most recently 't Hooft has advocated\tHooftEK\ that these interactions are important at the caustic origin of the horizon.) While this is the correct description of a test particle in the background of a gravitational shock wave like Aichelburg-Sexl\refs{\AichelburgDH}, when {\it two} such shocks interact the result is black hole formation\EardleyRE, and this black hole grows with center of mass energy.  We believe that this is the relevant physics here.  In this situation the postulate that the number of degrees of freedom of a black hole is determined by the Bekenstein-Hawking entropy
could lead to a self-consistent reduction of the number of degrees of freedom.  

Commutativity was revisited in the context of string theory in\refs{\LoweAC,\PolchinskiTA}, where the authors argued for a non-vanishing commutator such as \bhcom, though without unanimity on the presence of a gauge invariant signal.  Moreover, these authors argued that the relevant intermediate configuration producing the nonlocality is a long string stretching from $x$ to $y$.  From our discussion, this is rather difficult to understand, since in general there is insufficient energy to produce such a long string; one would instead expect it to be produced at the analog of the time ${\hat T}'$.  It may be that upon closer examination the effects they calculate are small; further investigation of the relation between their arguments and ours are definitely merited.  

More recently, Banks and Fischler\BanksYP\ have argued for an explanation of horizon complementarity stemming from the failure of different time evolution operators in gravity to commute.  These arguments do not superficially connect with those based on the locality bound, but perhaps a deeper relation exists.

We have attempted to make a sharp statement of the role of gravitational non-localities in decomposition of the Hilbert space in a black hole spacetime.  The generality of our arguments are worth emphasizing. Specifically, our argument for a flaw in Hawking's reasoning applies independent of whether or not string theory is the underlying theory of quantum gravity.
The reasoning presented in this paper just relies on generic features of gravitational physics.  The genericity of the argument against information loss is a satisfying mirror of the genericity of Hawking's arguments for black hole evaporation in an arbitrary quantum field theory coupled to gravity.  Of course, the restrictions placed on the underlying theory that result from demanding that it give a self-consistent description with a reduced number of degrees of freedom could be quite severe and may lead to something more radical than string theory.  For an attempt to formulate such a theory, see \refs{\BanksYP,\BanksVP}.

Other approaches to the information paradox have been explored in work by Jacobson (see \refs{\JacobsonVX} and references therein) and Horowitz and Maldacena\refs{\HorowitzHE}.  Jacobson investigates the role of a hypothetical cutoff at the Planck scale.  Such a cutoff is, of course, frame dependent.  We feel that the elimination of degrees of freedom arising from strong gravitational effects is more subtle than this and cannot simply be summarized in such a cutoff.  Horowitz and Maldacena have recently proposed a resolution of the information paradox involving a final state boundary condition on the black hole singularity.  Since this approach effectively moves information backwards in time in the internal region, we expect it to conflict with usual expectations for observations made by inside observers.  This proposal relies on assuming independent inside and outside Hilbert spaces, an assumption which we have strongly questioned, but perhaps could be a relevant piece of the physics in the complementary description appropriate to an outside observer.

One might also ask, if Hawking's argument for information loss is undermined by strong gravitational backreaction, whether his argument that black holes evaporate is safe.  We believe that the answer is yes, and that strong gravitational backreaction is relevant to questions of independence of the external and internal Hilbert spaces, but not to the question of the existence of the Hawking radiation.  For example, in two-dimensional models\refs{\CallanRS}, one may argue for the existence of Hawking radiation solely from diffeomorphism invariance manifested as conservation of the stress tensor, together with the existence of the conformal anomaly\refs{\ChristensenJC}.  This argument doesn't appear to refer to ultra-high energies and may reflect generalities of  the higher dimensional case.    Thus, our point of view is that strong gravitational effects are relevant when questions about localizing information are asked, but not when more coarse-grained features such as the existence of a Hawking flux are probed.

Finally, we comment on the role of two-dimensional models. In two dimensions gravity has no local dynamics, though black hole formation and evaporation can be described\refs{\CallanRS}.  Therefore our arguments regarding the locality bound being connected to strong gravitational backreaction should be reexamined in this context.  One alternative is that the two-dimensional story is simply different; for example, perhaps a black hole remnant containing the incident information {\it is} left behind.  In two dimensions this does not run into the obvious problem of infinite remnant production endemic to higher-dimensions.  Or, perhaps an underlying consistent description of two-dimensional gravity should involve strings, in which case we may find an analog to our locality bound arguments with string creation replacing strong gravitational effects.

\newsec{Complementarity for general horizons}

One obvious question is whether horizon complementarity holds in cosmological contexts, specifically in de Sitter space, as suggested in \refs{\Bankslittle,\BoussoNF,\DysonNT}.  If so, then the entire cosmology of de Sitter space might be described in terms of a finite number of degrees of freedom in a causal patch\refs{\Bankslittle,\Fischler}.  It is certainly possible that de Sitter space does not exist as a stable solution of a fundamental theory of quantum gravity; known string theory constructions\refs{\KachruAW} are unstable, and a more general argument\refs{\GiddingsZW} indicates that in any  theory  a de Sitter vacuum with compactified extra dimensions has a generic instability.  Nonetheless, one can investigate some aspects of this question.

Indeed, aspects of our analysis extend to more general horizons. Hawking radiation is generally present in spacetimes whose near-horizon geometry is that of Rindler space.\foot{Not all horizons are of this type, of course.  Extremal Reissner-Nordstrom black holes and D3-branes, for example, have AdS near-horizon geometries and do not radiate.}  A Rindler observer detects a thermal spectrum simply as a consequence of the Unruh effect.  While this may not always lead to an information paradox, generalized horizon complementarity suggests that degrees of freedom on opposite sides of the horizon are not independent and each observer has access to a complete description of the physics.

In particular, the observer-dependent horizon in de Sitter space is of this Rindler type.  The static patch metric,
\eqn\dsmetric{ds^2 =  -(1-{r^2 \over R^2})dt^2 + (1-{r^2 \over R^2})^{-1} dr^2 + r^2 d\Omega^2}
with dS radius $R = \sqrt{3 \over \Lambda}$, becomes the Rindler metric \rindler\ in the limit $r \to R$ with the coordinates $\theta = t/R$ and $\rho = \sqrt{\pm2R(r-R)}$.  Hawking radiation is indeed emitted from the horizon \refs{\GibbonsMU}, and, as a result, a static observer measures a thermal bath of temperature $T=\frac{1}{2\pi R}$.

We can apply the previous black hole locality bound analysis to dS merely by changing the length scale with the substitution $4M \to R$.  Again, we can conclude that QFT in a background, here dS, overcounts the number of independent degrees of freedom, and we may not treat operators on either side of the horizon as acting on separate Hilbert spaces.  This suggests a justification for the viewpoint of 
\refs{\Bankslittle,\BoussoNF,\DysonNT} that the observables in each static patch represent a complete basis of the total Hilbert space.  

One notable difference, however, is that, unlike an evaporating black hole, which has only outgoing radiation, dS is in thermal equilibrium, with radiation being both emitted and absorbed.  Because the static patch has finite volume, there are fundamental limits on the amount of information an observer can gather.  In particular, the N and D bounds \refs{\BoussoNF} imply that at most half the entropy is ever available to an observer, and \refs{\Page} showed that at least half the entropy is needed to extract information from the Hawking radiation.  Because the upper bound on the information retention time is infinite, eternal dS is not subject to an information paradox.\foot{Assuming information could somehow be retrieved from the Hawking radiation, \refs{\DanielssonTD} estimated the retrieval time to be $\tau_r \sim R^3$, similar to that of a black hole.   However, \refs{\DanielssonWB} argued that the thermalization time is also $\tau_{th}\sim R^3$, so a paradox is still avoided}  So the primary role of the locality bound and complementarity in the cosmological context is apparently to constrain and relate the fundamental degrees of freedom.

\newsec{Conclusion}

Hawking's derivation of information loss in black hole evaporation relies on the statement that one can decompose the Hilbert space into independent internal and external Hilbert spaces.  This paper has argued that a derivation of this statement from quantum field theory in a semiclassical background fails to be self-consistent:  when addressing the question of independence of inside and outside information, strong gravitational effects become relevant and the calculation breaks down.  We suggest that this is the loophole in Hawking's original argument that evades the 
black hole information paradox, while providing further support for the idea of black hole complementarity.

Our arguments are apparently independent of the underlying dynamics of gravity, though may ultimately provide clues to its nature.  We believe that our statement, via the locality bound, about when locality breaks down in physics may be part of a deeper self-consistent structure of quantum gravity.  The obvious outstanding problem is to find the relevant fundamental laws, whether non-perturbative string/M theory or something else.  (For one attempt at an alternative formulation, see \refs{\BanksYP,\BanksVP}.) We also believe that our arguments based on the locality bound begin to supply a piece of the framework for understanding black hole complementarity, or more generally horizon complementarity. 

Another outstanding question is to find a more general statement of the locality bound.  For example, we might state a weak version of the gravitational locality bound as follows:  

{\narrower\smallskip\noindent {\it Gravitational locality bound - weak version}: we cannot rely on calculations of local quantum field theory when discussing operators corresponding to degrees of freedom within a mutually created trapped surface. \smallskip}

%

\noindent More generally, one might not insist on trapped surface formation, but instead expect locality to fail simply when perturbations of the geometry/causal structure become large.  Moreover, in a specific theory of quantum gravity such as string theory, other effects may lead to locality violation even sooner, for example through creation of long strings.  Such modifications of the locality bound were briefly discussed in \GiddingsPT.

It's difficult to formulate a more specific criterion even for the weak gravitational bound.  For example, in a general curved background, it is very difficult to find a criterion for two high energy particles -- or more general degrees of freedom -- to form a black hole.  In a flat background we've used such a criterion \lbcrit, namely that in the center of mass frame a closed trapped surface should form when the particles reach a separation comparable to the Schwarzschild radius determined by their center of mass energy.  Such a criterion can be rigorously justified\EardleyRE.  We need  a more general statement.  We can seek counsel from the masters:  Thorne's hoop conjecture\refs{\Thorne} states: ``Horizons form when and only when a mass $M$ gets compacted into a region whose circumference in EVERY direction is $C\roughly< 4\pi GM/c^2$."  However, this does not provide sufficient guidance:  in a general curved spacetime we do not know how to define the mass in a region, and moreover the size of the region depends on choices such as that of a spacelike slice on which the size is measured.  
An interesting problem is to try to arrive at more general and correct criteria.

\bigskip\bigskip\centerline{{\bf Acknowledgments}}\nobreak

We wish to thank T. Banks, T. Erler, B. Freivogel, J. Hartle, S. Hawking, D. Marolf, A. Strominger, L. Susskind, and B. Williams for valuable discussions.  Parts of this work were carried out while SG was a guest of the Mitchell Institute, Texas A\&M, where a preliminary version of this work was presented at the conference ``Holography and the AdS/CFT correspondence," at the Superstring Cosmology workshop, Kavli Institute for Theoretical Physics, and at the Aspen Center for Physics.
SG gratefully acknowledges the support of both the Mitchell Insitute and the KITP.  This work was also supported in part by DOE grant DE-FG02-91ER40618.

\appendix{A} {Wavepacket propagation}
We wish to express a gaussian wavepacket at time $T'$, \gaussop,  in terms of operators at an earlier time $\hatT'$.  Using Green's theorem we write $\phi(x)$ in terms of data on an earlier slice:
\eqn\greenstheorem{\phi(x) = -\int d^3{\bf \hatx} \left\{ \phi(\hatx)  \partial_{\hat t} G_R(x, \hatx) - {\dot\phi(\hatx)}  G_R(x, \hatx) \right\}\ .}
Here we use the operator equation of motion, $\sq\phi=0$, and the 
free retarded Green's function is simply
\eqn\greensfunction{G_R(x,\hatx) = i\int \frac{d^3\bk}{(2\pi)^32\omega_k}  \left( e^{ik(x-\hatx)} - e^{-ik(x-\hatx)} \right)\ .}
Combining \greenstheorem\ and \gaussop\ , we have the wavepacket operator written in terms of operators at the earlier time,
\eqn\backprop{\phi_\delta(\bx,\bp,T') = -\int d^3 \bx' d^3 \hatbx {e^{-(\bx'-\bx)^2/2\delta^2 - i\bp\cdot(\bx' - \bx)} \over (\sqrt{2\pi} \delta)^d }\left\{ \phi(\hatx)  \partial_{\hat t} G_R(x', \hatx) - {\dot\phi(\hatx)}  G_R(x', \hatx) \right\}\ }
where $\hatx=(\hatbx, {\hat T}')$.
We perform the gaussian integral over $\bx'$ exactly to yield a gaussian integral in the momentum $\bk$.  Because the mean wavepacket momentum $|\bp|$ is much larger than the spread $\delta^{-1}$, we approximate this integral by neglecting terms of order $\frac{1}{|\bp|\delta}$.  The wavepacket propagates approximately classically, with the position of the peak following the path $\bx(t) = \bx - (T'-t) {\hat {\bf e}}_\bp$.  The resulting expression gives  $\phi_\delta(\bx,\bp,T')$ in terms of operators at the earlier time $\hatT'$
\eqn\wpatthat{\phi_\delta(\bx,\bp,T') \sim \int d^3 \hatbx  \left\{ Re f(\bx(\hatT'), \hatbx) \phi(\hatx)   -  \frac{{Im f(\bx(\hatT'), \hatbx)}}{\omega_{\bp}} \dot\phi(\hatx) \right\} }
where the function $f(\bx(\hatT'), \hatbx)$ is given by
\eqn\function{f(\bx(\hatT'), \hatbx) = e^{-(\hatbx-\bx(\hatT'))^2/2\delta^2 - i\bp\cdot (\hatbx-\bx(\hatT'))} \ .}
We now recognize the RHS of \wpatthat\ as a modified wavepacket, similar in form but slightly more complicated than \gaussop, with position $\bx(\hatT')$ and momentum $\bp$, and which we denote ${\tilde \phi_\delta}(\bx(\hatT'),\bp,\hatT')$.  Propagating ${\dot \phi_\delta}(\by,\bk,T')$ back to $\hatT'$, using an almost identical calculation, gives a similarly modified wavepacket ${\tilde {\dot \phi_\delta}}(\by(\hatT'),\bk,\hatT')$.
Having demonstrated that the wavepackets propagate back in time roughly classically, we can rewrite the \commreln\ as simply
\eqn\commpropback{[\phi_\delta(\bx,\bp,T'),{\dot \phi}_\delta(\by,\bq,T')] \sim [{\tilde \phi}_\delta(\bx(\hatT'),\bp,\hatT'),{\tilde {\dot \phi}}_\delta(\by(\hatT'),\bq,\hatT')] \ .}

\listrefs
\end